\documentclass[12pt,letterpaper]{article}
\usepackage[utf8]{inputenc}

 \pdfoutput=1 
\usepackage{color}
\definecolor{darkblue}{rgb}{0.1,0.1,.7}
\usepackage[colorlinks, linkcolor=darkblue, citecolor=darkblue, urlcolor=darkblue, linktocpage]{hyperref}
\usepackage[]{amsmath}
\usepackage[]{graphicx}
\usepackage[]{latexsym}
\usepackage[utf8]{inputenc}
\usepackage{lipsum}
\usepackage{slashed,graphicx,color,amsmath,amssymb}
\usepackage[mathscr]{eucal}
\usepackage{mathrsfs}
\usepackage[margin=10pt,font=small,labelfont=bf]{caption}
\usepackage{amscd}
\usepackage{bm}
\usepackage{xcolor}
\usepackage{upgreek}
\usepackage[square, comma, sort&compress,numbers]{natbib}
\usepackage[all,cmtip]{xy}
\usepackage[margin=1.8in]{geometry}
\usepackage{cleveref}
\usepackage[color=cyan!30!white,linecolor=red,textsize=footnotesize]{todonotes}
\usepackage{soul}
\usepackage{dsfont}
\numberwithin{equation}{section}
\hypersetup{
	pdfstartview={FitH},    
	pdftitle={The Schwarzian sector of higher spin CFTs},    
	pdfauthor={Datta},     
	colorlinks=true,       
	linkcolor=blue,          
	citecolor=blue!59!black! ,        
	filecolor=magenta,      
	urlcolor=blue           
}

\newcommand{\tr}{\mathrm{Tr}\,}

\def\btau{{\bar{\tau}}}

\def\bq{{\bar{q}}}

\def\bh{{\bar{h}}}

\def\cN{\mathcal{N}}

\def\ie{{\it i.e.~}}
\def\eg{{\it e.g.~}}

\def\nn{\nonumber}

\def\l1{{{1-loop}}}

\def\by{{\bar{y}}}

\def\n1{\Bigg|_{n=1}}

\def\n{{(n)}}
\def\tr{{Tr}}

\def\cN{\mathcal{N}}

\def\tr{\text{Tr}}

\def\bL{\bar{L}}

\def\bq{\bar{q}}
\def\bnu{\bar{\nu}}

\def\bh{{\bar{h}}}

\def\beq{\begin{equation}}
\def\eeq{\end{equation}}

\def\bea{\begin{eqnarray}}
\def\eea{\end{eqnarray}}
\def\nn{\nonumber}

\def\l1{{\text{1-loop}}}

\def\by{{\bar{y}}}

\def\n1{\Bigg|_{n=1}}

\def\n{{(n)}}
\def\tr{\text{Tr}}

\def\cN{\mathcal{N}}

\def\ket#1{|#1\rangle}

\def\vev#1{\langle{#1}\rangle}

\def\cW{\mathcal{W}}

\def\be{\begin{equation}}
\def\ee{\end{equation}}
\def\bal{\begin{array}{l}}
\def\ba#1{\begin{array}{#1}}  
	\def\ea{\end{array}}
\def\bea{\begin{eqnarray}}
\def\eea{\end{eqnarray}}
\def\beas{\begin{eqnarray*}}
	\def\eeas{\end{eqnarray*}}

\def\nn{\\\nonumber}

\def\vev#1{\langle #1 \rangle}

\def\nn{\nonumber}
\def\bit{\begin{item}}
	\def\eit{\end{item}}
\def\benu{\begin{enumerate}}
	\def\eenu{\end{enumerate}}
\def\tr{{\rm tr}}

\def\sln{\mathfrak{sl}(N)}
\def\Wn{\mathcal{W}_N}

 \makeatletter
 \g@addto@macro\bfseries{\boldmath}
 \makeatother

\def\adsthreecfttwo{\text{AdS$_3$/CFT$_2$}}

\def\adstwo{\text{AdS$_2$}~}

\def\id{{\mathds{1}}}

\makeatletter
\if@todonotes@disabled

\else

\fi
\makeatother

\begin{document}

\definecolor{tinge}{RGB}{255, 244, 195}
\sethlcolor{tinge}
\setstcolor{red}

\vspace*{-.5in} \thispagestyle{empty}
\begin{flushright}
\texttt{CERN-TH-2021-009}
\end{flushright}
\vspace{.2in} {\Large
\begin{center}
{\Large \bf
The Schwarzian sector of higher spin CFTs}
\end{center}}
\vspace{.2in}
\begin{center}
{Shouvik Datta}
\\

\vspace{.4cm}
\textit{\small Department of Theoretical Physics, CERN,\\
	1 Esplanade des Particules, Geneva 23, CH-1211, Switzerland.}
\\ \vspace{.2cm}
\par \begingroup\ttfamily\small
sdatta@cern.ch\par
\endgroup


\end{center}

\vspace{.5in}

\begin{abstract}
\normalsize{Two-dimensional conformal field theories with Virasoro symmetry generically contain a Schwarzian sector. This sector is related to the near-horizon region of the near-extremal BTZ black hole in the holographic dual. In this work we generalize this picture to CFTs with higher spin conserved currents. It is shown that the  partition function in the near-extremal limit  agrees with that of BF higher spin gravity in AdS$_2$ which is described by a generalized Schwarzian theory. We also provide a spectral decomposition of Schwarzian partition functions via the $\mathcal{W}_N$ fusion kernel and consider supersymmetric generalizations. }
\end{abstract}

\vskip 1cm \hspace{0.7cm}

\newpage

\setcounter{page}{1}

\noindent\rule{\textwidth}{.1pt}\vspace{-1.2cm}
\begingroup
\hypersetup{linkcolor=black}
\tableofcontents
\endgroup
\noindent\rule{\textwidth}{.2pt}

\section{Introduction}
The near-horizon geometry of extremal black holes universally contain an AdS$_2$ throat. This feature plays a crucial role in the microscopic understanding of black hole entropy in string theory \cite{Sen:2008vm,Mandal:2010cj,Sen:2014aja}. In the recent years, there has   been an upheaval of interest in the AdS$_2$/CFT$_1$ correspondence that has revealed the chaotic nature of black holes \cite{Cotler:2016fpe,Jensen:2016pah} and the unitarity of black hole evaporation \cite{Almheiri:2019qdq,Penington:2019kki}. A major player in this development has been the SYK model, whose soft mode is described by the Schwarzian theory -- with the action being the Schwarzian derivative of time reparametrizations \cite{Maldacena:2016hyu,Sarosi:2017ykf}. This quantum mechanical theory is also equivalent to JT gravity on a disk \cite{Maldacena:2016upp}. This lower dimensional avatar of holography has woven rich connections between black holes, matrix models, topological recursion and non-perturbative physics \cite{SSS,Okuyama:2019xbv}. 

It is worthwhile to explore general situations where the dynamics of the Schwarzian theory can universally emerge. The main reason being that the Schwarzian arises in the \textit{near}-AdS$_2$ regime. This is a departure from pure-AdS$_2$, where   reparametrization symmetry is broken, and the Schwarzian emerges as a Nambu-Goldstone mode \cite{Maldacena:2016upp}.
In the context of \adsthreecfttwo, the \adstwo throat develops in the near-horizon region of BTZ black holes close to extremality and the manner in which the Schwarzian is embedded was concretely shown in \cite{Ghosh:2019rcj} (see also \cite{Cotler:2018zff}). In fact, it was demonstrated that this feature does not even require the 2d CFT to be holographic and holds more generally for $c>1$ theories with Virasoro symmetry that have a twist gap. 
The goal of this work is to consider CFTs with larger symmetry algebras. Specifically, we have in mind theories with higher spin conserved currents, or $\mathcal{W}$-algebras instead of Virasoro. These CFTs are dual to Vasiliev-like theories in AdS$_3$, which describe a sector of string theory in the tensionless limit \cite{Gaberdiel:2010pz,Gaberdiel:2012uj}. A natural question is then how the Schwarzian dynamics gets adapted to this setting. Addressing this toy problem would orient us to eventually understand analogous structures in full-fledged string theoretic realizations of \adsthreecfttwo\,\cite{David:2002wn,Eberhardt:2019ywk}.\footnote{For a different perspective, see \cite{Maity} for the relation of the Schwarzian of covering maps to a class of correlators of the symmetric orbifold.}

The exact gravity dual of the SYK model still eludes us. There are indications that bulk theory has an infinite number of local fields \cite{Gross:2017hcz}. Furthermore, since the theory lacks a gap in its spectrum and owing to bounds on the branching time, sub-AdS locality might be absent \cite{Zhang:2020jhn}. These two features bear resemblance to higher spin gravity theories in AdS.  In \cite{Gonzalez:2018enk} higher spin gravity in \adstwo has been studied, with a view towards constructing the holographic dual of a SYK model with additional symmetries. It incorporated the BF formalism for 2d topological gravity \cite{Alkalaev:2013fsa} and showed its relation to the higher spin generalization of the Schwarzian theory. In particular, the logarithmic corrections to the entropy (or one-loop corrections to the partition function) were found to depend on the number of higher-spin fields present. As a part of this work, we precisely recover this partition function from an \adsthreecfttwo~perspective.

We shall be considering $\mathcal{W}_N$ CFTs in the irrational regime $c>N-1$. These are slightly different from the coset-models appearing in the Gaberdiel-Gopakumar duality \cite{Gaberdiel:2010pz}. These theories are special in the sense that unitarity enforces a lower bound on the conformal dimension that scales linearly with the central charge \cite{Afkhami-Jeddi:2017idc}. This implies at large $c$, all non-trivial primaries are heavy states. The holographic dual for these irrational CFTs would correspond to theories of pure higher spin gravity in AdS$_3$. Some recent progress in this direction has been made in \cite{Alday:2020qkm}. The BTZ black hole still continues to be a solution of this higher spin theory. We study the partition function of the graviton and higher-spin fields in the BTZ background and focus on the near-extremal limit. This provides a generalization of the analysis performed in \cite{Ghosh:2019rcj} and shows the emergence of a sector described by higher-spin Schwarzian theories. We also consider higher spin $\cN=1$ and $\cN=2$ supergravities, and find that  analogous super-Schwarzian sectors exist for some specific choice of spin-structures.

Upon obtaining the Schwarzian partition functions, we explore how the states of theory are distributed. This can be extracted directly from the partition function via the usual method of inverse Laplace transforms. One can adapt a slightly different point of view, inspired by Toda theories. The primary states in the $\mathcal{W}_N$ CFTs can be parametrized by momentum vectors that live in the root lattice of $\sln$. We can see how states are distributed in this lattice and how the underlying symmetries play a role. This objective is efficiently achieved by utilizing the fusion kernel of $\mathcal{W}_N$ CFTs. The fusion kernel was found earlier in the context of studying defects in \cite{Drukker:2010jp}. Using this, we find a simple result for the distribution function in terms of the Cartan matrix of $\sln$. 

There exist more general black hole solutions in 3d higher spin gravity \cite{Gutperle:2011kf,Ammon:2012wc}. These are generalizations of the BTZ black holes which have non-vanishing higher spin hair. The notion of extremality for these black holes can be phrased in a suitable gauge invariant manner using the Chern-Simons formulation of higher spin (super)gravity \cite{Datta:2012km,Datta:2013qja,Banados:2015tft,Henneaux:2015ywa}. It is an interesting question what is the analogue of the Schwarzian theory in the extremal limit of these black holes. This requires knowledge of one-loop partition functions in these black hole backgrounds. Unfortunately, these are currently not known and the modular properties of these partition functions are known only perturbatively in some special cases \cite{Datta:2014zpa,Iles:2014gra}. We do not address this question in this paper, hoping to return to this interesting direction in the near future.

This paper is organized as follows.  In Section \ref{sec:extremal-limit} we study the extremal limit of the partition function of higher spin (super)gravity in the BTZ black hole background. The resulting partition functions are shown to agree with that of BF higher spin gravity in \adstwo or higher spin Schwarzian theories.  The spectral decomposition is investigated in Section \ref{sec:spectra}, using both the inverse Laplace transform and the fusion kernel. We conclude in Section \ref{sec:conclusions}. Appendix \ref{app:sln} recollects some details on the representation theory $\sln$ that play an important role in finding the spectral decomposition.

\section{The extremal limit}
\label{sec:extremal-limit}

In this section we consider partition functions of higher spin fields in the near-extremal BTZ black hole. A convenient formulation of higher spin fields in 3d is by using Chern-Simons theory. We consider higher spin theories of (super)gravity where the SL(2,$\mathbb{R}$), corresponding to the pure gravity sector, is principally embedded in the higher spin algebra. In what follows, we shall employ one-loop determinants of higher spin fields in thermal AdS$_3$. These can be obtained by integrating the heat kernel which is known for arbitrary spin-$s$ fields \cite{David:2009xg}. As shown in \cite{Gaberdiel:2010ar}, the partition functions on thermal AdS$_3$ with an appropriate field content agree with vacuum characters of $\mathcal{W}$-algebras. 

We consider the BTZ black hole as a solution of higher spin (super)gravity in AdS$_3$. As alluded to in the introduction, there are certainly more general black holes in these theories of gravity. However, we stick to the simplest case where the higher spin charges and chemical potentials are turned off. The Euclidean BTZ has a topology of a solid torus and we are interested in the partition function as a function of modular parameters of the boundary torus
\begin{align}\label{zBTZ00}
Z_{\rm BTZ}(\tau,\btau)  = \tr  \left[q^{L_0-\frac{c}{24}} \bar{q}^{\bar{L}_0-\frac{c}{24}}\right], \qquad q=e^{2\pi i\tau}, ~ \bq=e^{2\pi i\btau}~. 
\end{align}
The above quantity includes a classical contribution of the BTZ black hole from its on-shell action and fluctuations of higher spin field about this background, encapsulated through the one-loop determinants. 
The modular parameters are chosen to be\footnote{We have set the circumference  of the spatial circle of the torus to be $L=2\pi$. }
\begin{align}
	\tau = \frac{\theta + i\beta}{2\pi}\, , \quad \btau = \frac{\theta - i\beta}{2\pi}~. 
\end{align}
Here, $\theta$ is the chemical potential conjugate to the angular momentum, $J = L_0 - \bL_0$. 
We shall often find it convenient to work with  left/right moving temperatures related to the above by 
$\tau=i\frac{\beta_L}{2\pi}$ and $\btau=-i\frac{\beta_R}{2\pi}$.  

The partition function \eqref{zBTZ00} can be obtained from that of thermal AdS$_3$ by a S-modular transformation. Just like the case of pure gravity, it is one-loop exact and encodes excitations of the boundary gravitons and higher-spin fields. It takes the following form
\begin{align}\label{zbtz-hs}
	Z_{\rm BTZ}(\tau,\btau) = \chi_\id (-1/\tau) \chi_\id (1/\btau)~,
\end{align}
where, $\chi_\id (\tau)$ is the vacuum character of the specific $\mathcal{W}$-algebra in question. The above relation also reflects the asymptotic symmetries of the higher spin gravity theory. To illustrate the near-extremal limit let us recall the standard relation between the left/right-moving temperatures and the horizon radii of the BTZ black hole
\begin{align}
\beta_L ^{-1} = r_+ - r_-~, \qquad \beta_R ^{-1} = r_+ + r_-~,
\end{align}
in units of $\ell_{\rm AdS}=1$. The Hawking temperature ($T_H=1/\beta$) is given by 
\begin{align}
	\beta = \frac{\beta_L+\beta_R}{2} =  \frac{2r_+}{r_+^2 -r_-^2}~. 
\end{align}
The geometrical description of the horizon is no longer valid once higher spin fields are present. Thermodynamics of these objects are formulated using gauge-invariant notions, such as holonomies and Wilson loops. The two equations above are merely used to identify the regime of CFT temperatures in the extremal limit, and this is not different from Einstein gravity as we are considering the black hole without higher spin charges. 

 For the black hole to be a dominant saddle we also require $\beta_L\beta_R \leq (2\pi)^2$. We consider the holographic dual to a theory with large central charge. The central charge, in turn, sets the temperature scales to recover the Schwarzian theory.  As shown for the pure gravity case in \cite{Ghosh:2019rcj}, this is the near-extremal BTZ ($r_+\to r_-$) and corresponds to the certain regime in parameter space in the grand canonical ensemble $(\beta,\theta)$. When additional fields sourcing gravity are present, the regime generalizes to 
\begin{align}\label{NEXTregime}
c\gg O(N), \qquad \beta_L\sim O(c), \qquad \beta_R \lesssim 1/c~.
\end{align} 
Here, $O(N)$ is the number of higher spins fields present in gravity, or the number of conserved currents in the CFT. 
Note that \eqref{NEXTregime} implies $\beta_L \approx 2\beta$ (and $\beta_R = \beta + i\theta$). The characters appearing in the partition function \eqref{zbtz-hs} are already S-modular transformed, indicating they are adapted for a high-temperature expansion. On the other hand, from \eqref{NEXTregime} we require a  low-temperature for the left-moving part. Therefore, we need to S-transform `back' the modular functions appearing to arrive at the $\beta_L \sim O(c)$ regime.

\subsection{Bosonic higher spins}

As our first example of a higher spin gravity theory we consider a truncated tower of bosonic higher spins \ie with fields having spins $s=2,3,\cdots, N$. This theory is described by a Chern-Simons theory based on the gauge group SL$(N,\mathbb{R})\times$ SL$(N,\mathbb{R})$. The partition function including the one-loop corrections is given by $|	\chi_\id(\tau)|^2$, where
\begin{align}\label{vac-char-WN}
	\chi_\id(\tau) = q^{-\frac{c}{24}}\prod_{s=2}^N \prod_{n=s}^\infty \frac{1}{1-q^n} =  q^{-\frac{c-(N-1)}{24}} \frac{\prod_{m=1}^{N}(1-q^m)^{N-m}}{\eta(q)^{N-1}} ~.
\end{align}
This is also the vacuum character of $\mathcal{W}_N$ CFTs with $c>N-1$. In other words, it counts the descendants of the identity which are of the form $(W^{(s)}_{-n_1})^{N_1} (W^{(p)}_{-n_2})^{N_2}\cdots \ket{0}$ (here, $W^{(s)}_{-n_i}$ indicates a generator of  spin $s$ and mode number $n_i$). The product in the numerator prevents overcounting of  the states that are equivalent to the vacuum. 
Using this expression we can find the partition function \eqref{zbtz-hs} in the regime \eqref{NEXTregime}.

Let us focus on the left-moving part first. For  $\beta_L \sim O(c) \gg O(N)$, the product appearing in the numerator of the character \eqref{vac-char-WN} and the Dedekind-etas are 
\begin{align}\label{S-duals}
	&\prod_{m=1}^{N} (1-e^{-\frac{(2\pi)^2 m }{\beta_L}})^{N-m}  \approx \prod_{m=1}^{N} \left(\frac{(2\pi)^2 m }{\beta_L}\right)^{N-m}= \mathcal{K}\left(\frac{(2\pi)^2 }{\beta_L}\right)^{\frac{N(N-1)}{2}} ,\nn \\
	&~\eta\left(e^{-\frac{(2\pi)^2 m }{\beta_L}}\right)^{N-1}\approx  \left(\frac{2\pi}{\beta_L}\right)^{-{(N-1)\over  2}} e^{-(N-1)\frac{ \beta_L }{24}} ~. 
\end{align}
where, $\mathcal{K}$ is a constant depending only on $N$.\footnote{This can be written explicitly using generalized Riemann-zeta functions and the Glaisher–Kinkelin constant. The details are however not important in the current context.} We have used the S-modular transformation of the Dedekind-eta function  $\eta(-1/\tau) = \sqrt{-i\tau}\,\eta(\tau)$. 
Combining the above two equations, the final result for the left moving character (at low temperatures) is
\begin{align}\label{chiL}
	\chi_\id \left(\frac{2\pi i}{\beta_L}\right) &\approx (2\pi)^{N(N-1)\over 2} {\left(\frac{2\pi}{\beta_L}\right)^{N^2-1\over 2}} \exp \left[(N-1)\frac{\beta_L}{24} +\frac{\pi^2  c }{6\beta_L}  \right] \times \text{const.}  
\end{align}
The most important feature of the above expression is the generalized  power-law behaviour in the prefactor, $\beta_L^{-({N^2-1})/{2}}$. Part of the expression is also of the same form as the partition function of the higher spin Schwarzian theory \cite{Gonzalez:2018enk} -- we shall return to this aspect in a moment. The right-moving character is 
\begin{align}\label{chiR}
	\chi_\id \left(\frac{2\pi i}{\beta_R}\right) &\approx  \exp \left[  \frac{\pi^2  c }{6\beta_R} \right] ~. 
\end{align}
The expressions \eqref{chiL} and \eqref{chiR} obviously reduce to the pure gravity case for $N=2$ \cite{Ghosh:2019rcj}.

In order to make the relationship to the Schwarzian theory transparent, we recast the above results in the canonical ensemble of spins following \cite{Ghosh:2019rcj}. This ensemble has the temperature $\beta$ and the angular momentum $J$ fixed. It can be obtained by the Fourier transform 
\begin{align}\label{fourier-trans}
	Z_\beta (J) = \int_{-\pi}^\pi \frac{d\theta}{2\pi} e^{i\theta J} Z_{\rm }(\beta,\theta)~. 
\end{align} 
The near-extremal regime in this ensemble is given by $c\gg O(N)$, $\beta \sim O(c)$ and $J\gg c$. We can use the relations $\beta_L \approx 2\beta$ and $\beta_R = \beta + i\theta$, the integral \eqref{fourier-trans} then effectively becomes an inverse Laplace transform in the right-moving temperature $\beta_R$. This can be evaluated in the saddle point approximation at large $J$, with the result
\begin{align}\label{Z-can}
	Z_\beta (J)  ~\propto~ \left(\frac{c }{6J^3}\right)^{1\over 4} \left(\frac{\pi}{\beta}\right)^{(N^2-1)\over 2} 
	\exp \left[ 2\pi \sqrt{\frac{c J}{6}} -\beta \left(J- \frac{N-1}{12}\right) + \frac{\pi^2 c  }{12\beta} \right]~. 
\end{align}
We use the definitions
\begin{align}
	S_{\rm BH} =2\pi\sqrt{ \frac{c J}{6}}, ~ E_{\rm BH} = \left(J- \frac{N-1}{12}\right)~. 
\end{align}
and write \eqref{Z-can} as
\begin{align}
	Z_\beta (J)  
	&~\propto~ c^{\frac{3-N^2}{2}}    
	\exp \left[ \left(S_{\rm BH} - \frac{3}{2} \log S_{\rm BH} \right)-\beta E_{\rm BH}  \right]\, Z_{\rm Sch}(\beta)~. 
\end{align}
Writing it in this form makes it manifest that in the near-extremal limit, the right moving sector encodes the degeneracy of extremal black hole microstates whilst the left moving sector contains the excitations described by the Schwarzian theory. 
Finally, the Schwarzian partition function is
\begin{align}\label{Z-Sch-WN}
	Z_{\rm Sch}(\beta) =\left(\frac{\pi^2 c}{\beta}\right)^{(N^2-1)\over 2} 
	\exp \left[  \frac{\pi^2 c  }{12\beta} \right]~. 
\end{align}
The exponent in the prefactor (which is the one-loop correction to the free energy) above is half of the number of generators of $\sln$, the wedge-subalgebra of $\mathcal{W}_N$. In \cite{Stanford:2017thb}, it was shown on very general grounds that this exponent is given by half of the number of  zero mode configurations of the quadratic part of the Schwarzian action --  we need to quotient out these while performing the path integral. For the standard Virasoro case, there are three  zero modes corresponding to the action of the generators $L_{0,\pm1}$. For the higher spin case, the generalized Schwarzian action is invariant under the actions of $W^{(s)}_{m}$ for $-(s-1)\leq m \leq (s-1)$ -- this accounts for the $(N^2-1)/2$ in the exponent above. 
Whenever fermionic zero modes are present, they contribute to the exponent with an opposite sign. We shall encounter this phenomenon in the supersymmetric cases to be considered next. 

The fact that the number of zero-mode configurations is $(N^2-1)$ can also be seen from finite temperature propagators of the Schwarzian modes \cite{Haehl:2018izb}. The quadratic part of the Schwarzian action can be written in terms of 2-point functions of the conserved currents. The propagators can then be shown to exhibit poles in the frequency plane and there are exactly $(N^2-1)$ number of poles (with degeneracies included). Furthermore, in the context of chaos $(N^2-1)$ is also the total number of `pole skippings' in the retarded Green's functions of all the conserved higher spin currents.

The result \eqref{Z-Sch-WN} agrees with the partition function of 2d higher spin BF theory (on the disk)  based on the group SL$(N,\mathbb{R})$ \cite{Alkalaev:2013fsa,Gonzalez:2018enk}. The boundary modes of this  gravity theory is given by higher spin Schwarzian actions, see \eg \cite{Li:2015osa}.  One drawback of our present analysis is that we have not considered the most general gravity background since the higher spin charges are turned off. The above formula would get modified in such a case and it would be interesting to see this in detail.

A natural question at this point is how general this analysis is and what are the additional  assumptions we implicitly made. Firstly, writing the BTZ partition function as \eqref{zbtz-hs} we have assumed vacuum dominance in the S-dual channel. This dominance is justified provided the low-lying spectrum is sparse; this is quantified by the presence of a twist gap in the spectrum. An estimate of the suppression from non-vacuum conformal families can be found from the ratio of the full partition function (expanded in the S-dual channel) with $Z_{\rm BTZ}$. If there's a single primary at the twist gap, the ratio is
\begin{align}
	\frac{Z(\beta_L,\beta_R)}{Z_{\rm BTZ}(\beta_L,\beta_R)} = 1 + O\left(\beta_L ^{\frac{N(N-1)}{2}}e^{-\frac{(2\pi)^2}{\beta_R}\bar{h}_{\rm gap}}\right)~. 
\end{align}
The second term is essentially the ratio of $\chi_{h}(2\pi i /\beta_L)\chi_{\bh}(2\pi i /\beta_R)$ with $Z_{\rm BTZ}$. The factor of $\beta_L^\#$ essentially arises from the product in the first line of \eqref{S-duals}. Therefore, vacuum dominance implies the following condition on $\beta_R$
\begin{align}\label{bound0}
\beta_R \ll \frac{2\bar{h}_{\rm gap}}{N(N-1)\log c}~. 
\end{align}
It was shown by imposing positivity of the Kac determinant that all  $c>N-1$ CFTs with $\mathcal{W}_N$ chiral algebras have a  lower bound in the conformal dimension  that scales linearly with $c$ \cite{Afkhami-Jeddi:2017idc}.  Therefore, for large $c$ these theories do not have light primaries and all non-vacuum primaries are heavy. This feature weakens the bound \eqref{bound0} or widens the allowed range of $\beta_R$.  For holographic theories, however, there is a large degeneracy of primaries at this energy scale coming from states with large spin.  These states  precisely correspond to large extremal BTZ black holes in the bulk. The regime of temperatures that where the black hole  dominates is $\beta_L \sim O(c)$, which is large, and $\beta_R\lesssim 1/c$ so that $\beta_L\beta_R \leq (2\pi)^2$.\footnote{If we focus only on the large $c$ regime, the analysis of \cite{Hartman:2014oaa} holds true even for CFTs with additional conserved currents.} It is worthwhile to remark at this point that it was crucial that we have worked in the irrational $\mathcal{W}_N$ CFTs and not the minimal-models originating from the coset construction. The $\mathcal{W}_\infty [\lambda]$ minimal models have a dense spectrum of light states that prohibit the Hawking-Page transition and the black holes are not expected to dominate the canonical ensemble at any regime of temperatures \cite{Banerjee:2012aj}. 

\subsection{Supersymmetric generalizations}

We now turn to supersymmetric generalizations of the analysis in the previous subsection. In particular, we analyse $\cN=1$ and $\cN=2$ higher spin supergravities and study the extremal limit of their partition functions in the BTZ background. 

\subsubsection*{$\cN=1$ higher spin supergravity}

Theories of $\cN=1$ higher spin supergravity can be formulated as Chern-Simons theories based on the supergroup OSp$(N|N-1)$ for odd $N$ and OSp$(N-1|N)$ for even $N$ \cite{Candu:2013uya}. The dual CFTs correspond to theories with $\cN=1$ super-$\mathcal{W}_N$ symmetries. The partition function including one-loop fluctuations is given by \eqref{zBTZ00}, where the character is 
\begin{align}\label{N1-z}
	\chi_\id(\tau) &= q^{-\frac{c}{24}} \prod_{s=2}^{N} \prod_{n=s}^{\infty} \frac{1-q^{n-1/2}}{1-q^n}
	= q^{-\frac{c-\frac{3}{2}(N-1)}{24}} \frac{\vartheta_4(q)^{(N-1)/2}}{\eta(q)^{3(N-1)/2}} \prod_{n=1}^{N-1} \left[1-q^n \over 1-q^{n-\frac{1}{2}}\right]^{N-n}. 
\end{align}
The choice of the spin-structure for the fermions is made as follows\footnote{We thank the anonymous referee for providing this explanation.}: 
in the black hole background, a natural choice of boundary conditions is that fermions are anti-periodic along the thermal cycle (NS) and
periodic along the spatial cycle (R). This implies that once we switch to the S-dual channel, for which \eqref{N1-z} applies, the fermions are rendered to be periodic along the thermal cycle. 
 We consider the BTZ partition function as \eqref{zbtz-hs}. The manipulations are similar to the bosonic case and we find that the partition function in the extremal limit has the following factor that encodes the super-Schwarzian for higher spins
\begin{align}
		Z_{\rm Sch}(\beta) =\left(\frac{\pi^2 c}{\beta}\right)^{(N-1)\over 2} 
	\exp \left[  \frac{\pi^2 c }{12\beta} \right]~. 
\end{align}
In this case the prefactor originates not from the product in  \eqref{N1-z}, but solely from the Dedekind eta and Jacobi theta function. The exponent here is precisely one-half of the difference between the bosonic generators minus the fermionic generators of $\mathfrak{osp}(N|N-1)$ or $\mathfrak{osp}(N-1|N)$. In these wedge sub-algebras of $\cN=1$ super-$\mathcal{W}_N$, each multiplet contains one bosonic generator more than the fermionic ones. As discussed while considering bosonic higher spins, the quantity $(N-1)$ is the difference between the number of bosonic and fermionic zero mode configurations that need to be quotiented out while performing the path integral of the Schwarzian theory. For the case of $N=2$, we get a factor of $\beta^{-1/2}$ and this agrees with previous analyses in the literature \cite{Stanford:2017thb,Mertens:2017mtv}.

\subsubsection*{$\cN=2$ higher spin supergravity}
In the same vein, the Schwarzian sector of $\cN=2$ higher spin supergravity can also be found. The corresponding gauge group for the Chern-Simons description is SL$(N|N-1)\times$SL$(N|N-1)$. Although the analysis is similar to the bosonic and $\cN=1$ cases but there are some interesting differences. We shall see this as we proceed. The BTZ partition function is once again given by \eqref{zBTZ00}, with $\chi_\id(\tau)$ as follows
\begin{align}
	\chi_\id(\tau) &= q^{-\frac{c}{24}} \prod_{s=2}^{N} \left(\prod_{n=s}^{\infty} \frac{1}{(1-q^n)}\right) \left(\prod_{n=s-1}^{\infty} \frac{1}{(1-q^n)}\right) 
	\left(\prod_{n=s}^{\infty}  {(1-q^{n-\frac{1}{2}})^2}\right)~. 
\end{align}
This is the vacuum character of the $\cN=2$ super-$\Wn$ algebra. The choice of spin-structure is same as the $\cN=1$ case. The above expression can be rearranged as 
\begin{align}		\label{char-N2Wn}
	\chi_\id(q) 
	&= q^{-\frac{c-3N}{24}} \frac{\vartheta_4(q)^{N-1}}{\eta(q)^{3(N-1)}}
	\prod_{n=1}^{N-1} 
	\left[1-q^n\over 1 - q^{n-\frac{1}{2}}\right]^{2(N-n)} \frac{1}{1-q^n} ~. 
\end{align}
Taking the extremal limit of $Z_{\rm BTZ}$ as before, we observe that the partition function of the Schwarzian sector in this case is given by 
\begin{align}
	Z_{\rm Sch}(\beta) =
	\exp \left[  \frac{\pi^2 c }{12\beta} \right]~. 
\end{align}
There are no one-loop corrections, due to an exact Bose-Fermi cancellation. Each multiplet of the $\mathfrak{sl}(N|N-1)$ algebra has the same number of fermionic and bosonic generators and this gives a vanishing exponent for the prefactor. Furthermore, note that the above formula has got no dependence on the number of higher spin fields present. 

However, this is not the full story. Algebras with $\cN=2$ supersymmetry have  a spectral flow automorphism. The net contribution to the Schwarzian sector should then include a sum over spectrally flowed representations. To realize this, let us consider the flavoured version of the partition function 
\begin{align}\label{zBTZ-flav}
	Z(\tau,\btau, \nu,\bar \nu)  = \tr  \left[q^{L_0-\frac{c}{24}} \bar{q}^{\bar{L}_0-\frac{c}{24}} y^{J_0}\by ^{\bar{J}_0}\right], \quad q=e^{2\pi i\tau}, ~ \bq=e^{2\pi i\btau},~  y=e^{2\pi i\nu},~ \by=e^{2\pi i\bar\nu}.
\end{align}
The bosonic subgroup of $\mathrm{SL}(N|N-1)$ is $\mathrm{SL}(N)\times \mathrm{SL}(N-1)\times \mathrm{U}(1)$. In the bulk, the  U$(1)$ Chern-Simons gauge field leads to  changes in the classical action of the BTZ \cite{Kraus:2006wn}
\begin{align}\label{n2-tree}
	\log Z_{\rm BTZ}|_{\rm tree} = \frac{\pi^2 c}{6\beta_L} + \frac{\pi^2 c}{6\beta_R} - \frac{2\pi k \nu^2}{\beta_L }- \frac{2\pi k \bar \nu^2}{\beta_R }~. 
\end{align}
Here, $k$ is the level of the $\mathfrak{u}(1)$ current algebra and for $\cN=2$ theories $k=c/3$. The one-loop partition function with chemical potentials is a generalization of \eqref{char-N2Wn}
\begin{align}\label{n2-one-loop}
	Z_{\rm BTZ}|_{\text{1-loop}}  
		&= q^{\frac{N}{8}} \frac{\vartheta_4(y,q)^{N-1}}{\eta(q)^{3(N-1)}}
	\prod_{n=1}^{N-1} 
	{	(1-q^n)^{2(N-n)-1} \over (1 -y^{-1} q^{n-\frac{1}{2}})^{N-n}(1 - y q^{n-\frac{1}{2}})^{N-n}   } \times \text{anti-hol} ~. 
\end{align}
with $q=e^{-2\pi i /\tau}$ and $y=e^{2\pi i \nu  /\tau}$. We can combine \eqref{n2-tree} and \eqref{n2-one-loop} and take the extremal limit keeping the chemical potentials $(\nu,\bnu)$ fixed. The partition function for the Schwarzian sector reads
\begin{align}
Z_{\rm Sch}(\beta) = \frac{\cos(\pi\nu)^{N-1}}{\left(\tfrac{1}{2}+\nu\right)_{N-1}\left(\tfrac{1}{2}-\nu\right)_{N-1}}	\exp \left[  \frac{\pi^2 c }{3\beta}\left(\tfrac{1}{2}+\nu\right)\left(\tfrac{1}{2}-\nu\right) \right]~. 
\end{align}
Here, $(\tfrac{1}{2}\pm\nu )_{N-1}$ are Pochhammer symbols that arise from the product in \eqref{n2-one-loop}. The cosine factor originates from the S-modular transformation, $\vartheta_4 \mapsto \vartheta_2$. 
We can now sum over spectrally flowed sectors. For the spectral flow parameter $\mu$, the conformal dimension and $U(1)$ charge change  as follows
\begin{align}
	h_\mu \mapsto h + \mu Q +\frac{\mu^2}{6} c, \qquad Q_\mu \mapsto Q+ \frac{c}{3}\mu~.
\end{align}
In order to preserve periodicity conditions of the fermions we shall sum only over integral spectral flows, $\mu =n \in \mathbb{Z}$. We also need to know how the S-transformed vacuum characters change under this flow
\begin{align}
	\chi_\id(-1/\tau,\nu/\tau) \mapsto q^{\frac{c}{6}n^2} y^{\frac{c}{3}n}\chi_\id  (-1/\tau,(\nu+n)/\tau)~, \qquad q=e^{-2\pi i/\tau},~ y = e^{2\pi i \nu/\tau}~. 
\end{align}
The offset in the chemical potential incorporates the shift in the non-zero modes of the stress tensor under spectral flow, $L_m \mapsto L_m + n J_m$. 
Using these characters, taking the extremal limit and then summing over $n$ yields the following result for the Schwarzian partition function 
\begin{align}\label{n2-result}
	Z_{\rm Sch}(\beta) = \sum_{n \in \mathbb{Z}} \frac{\cos(\pi(\nu+n))^{N-1}}{\left(\tfrac{1}{2}+\nu+n\right)_{N-1}\left(\tfrac{1}{2}-\nu-n\right)_{N-1}}	\exp \left[  \frac{\pi^2 c }{3\beta}\left(\tfrac{1}{2}+\nu+n\right)\left(\tfrac{1}{2}-\nu-n\right) \right]~. 
\end{align}
This result agrees with \cite{GaiottoSachdev,Stanford:2017thb,Mertens:2017mtv} for $N=2$, which is the case of $\cN=2$ super-Virasoro for the 2d CFT. In the low energy limit of the $\cN=2$ SYK model, the above sum takes into account all possible windings of the axionic field, that controls the U$(1)$ rotations of the R-symmetry. 

For the $\cN=2$ super-Schwarzian theories in general, there is a discrete parameter $\hat q$ that serves as the compactification radius for the axionic field above \cite{GaiottoSachdev,Stanford:2017thb}.   In supersymmetric SYK models, this parameter encodes the number of fermions appearing in the $\cN=2$ supercharge. Upon contrasting the general expression for the Schwarzian partition function for arbitrary $\hat q$ with \eqref{n2-result}, we see that $\hat q=1$.\footnote{The reason why this happens can be understood from the realization of the ${\cal  N} = 2$ super-Virasoro (or super-$\cal W$) algebra using free bosons and fermions. The  spin-3/2 supercurrents $G^\pm$ are linear in the free fermions and also linear in the free boson, $G^{\pm}_n \sim \sum_{m}\alpha_{n-m}\psi_m$. In other words,  $G^\pm$ is of degree $1$ in the fermions which implies $\hat q=1$. Although the free field realization does not apply to irrational CFTs, it leads to the same value of the parameter $\hat q$. The spin-3/2 supercurrent corresponds to the gravitino in bulk dual and this information enters via the partition function \eqref{n2-one-loop}. From this perspective and from the final expression for $Z_{\rm Sch}(\beta)$ in \eqref{n2-result}, the choice $\hat q=1$ is naturally picked for the near-extremal BTZ black hole.}

\section{Spectral decompositions}
\label{sec:spectra}

With the results for the partition functions of the higher spin Schwarzian in hand, we study the density of states in these theories. We shall do this in two ways. The first method directly extracts the spectral density by an inverse Laplace transform of the partition function. The second method will make use of the fusion kernel of the $\Wn$ algebra. We shall focus only on  bosonic higher spins. The analysis for the supersymmetric cases can be performed in a similar manner. 

\subsection{Direct calculation of the density of states}
The partition function of the Schwarzian theory can be written as the following integral over the density of states. 
\begin{align}\label{dos1}
	Z_{\rm Sch}(\beta) = \int_0^\infty \rho(E) e^{-\beta E} dE~. 
\end{align}
Using \eqref{Z-Sch-WN} as $Z_{\rm Sch}(\beta)$, $\rho(E)$ can be obtained via an inverse Laplace transform. 
\begin{align}
	\rho(E) = \oint d\beta\, e^{\beta E}\, Z_{\rm Sch}(\beta) ~. 
\end{align}
This integral can be performed exactly with the result  
\begin{align}\label{dos2}
	\rho(E)  ~\propto~ c^{\nu+1} E^{\nu} \,I_{2\nu} \left(\pi\sqrt{\frac{c E}{3}}\right), \qquad \nu = \frac{N^2-3}{4}\,,
\end{align}
with $I_\nu(x)$ being the modified Bessel function  of the first kind. For $N=2$ this becomes
\begin{align}\label{N2-DoS}
	\rho(E)  ~\propto~ c\,  {\sinh\left(\pi\sqrt{cE\over 3}\right)} ~,
\end{align}
which agrees with the result of \cite{Stanford:2017thb,SSS} (to adapt the conventions  of \cite{SSS}, we can set $c=24\gamma$).  As emphasised in \cite{Charles:2019tiu,Maldacena:2016hyu}, at high energies the inverse Laplace transform can be performed by using a saddle-point approximation and quadratic fluctuations about the saddle precisely cancel with the $\beta^{-3/2}$ prefactor. This implies that there aren't any logarithmic corrections to the microcanonical entropy for the pure gravity ($N=2$) case. This phenomenon, however, ceases to happen for higher spins $N\geq 3$. 

\subsection{Decomposition from the fusion kernel}
\label{sec:fusion}
We shall now take a different route to obtain the spectral decomposition of the Schwarzian partition functions. The key concept which will be invoked is that of the fusion kernel for characters of $\Wn$ CFTs. Since we are in the irrational regime, we can parametrize the CFT data akin to that of Toda theories. These theories are generalizations of Liouville theory and contain higher spin conserved currents.

The central charge is parametrized using the background charge 
\begin{align}
	c = (N-1) + 12 \vev{\hat Q,\hat Q}~, \qquad \hat Q = (b+b^{-1}) \hat \rho~. 
\end{align}
Here, $\hat \rho$ is the Weyl vector  and $\vev{\cdot,\cdot}$ is the Cartan-Killing form of $\sln$. The conformal dimensions are given in terms of momentum vectors that live in the root space of $\sln$. 
\begin{align}\label{h-alpha-rel}
	h-\frac{c}{24} = - \frac{1}{2} \vev{\hat a,\hat a}~, \qquad \hat a = \hat Q - \hat \alpha~. 
\end{align}
Here, $\hat \alpha$ is an arbitrary vector in the root space of $\sln$ and can be written, for practical purposes, as a linear combination of the simple roots. For the case at hand we shall choose $\hat\alpha= - i \hat P$, (so that primaries are of the vertex operator form, $e^{-i\vev{\hat P, \phi}}$). 
The characters of non-vacuum primaries of $\Wn$ is 
\begin{align}
	\chi_\alpha(\tau) = \frac{q^{-\frac{1}{2}\vev{\hat a,\hat a}}}{\eta(\tau)^{N-1}}~. 
\end{align} 
Just like the Virasoro case, the S-modular transformation of the vacuum character of $\cW_N$ theory can also be written as a sum over all characters in the dual channel using the fusion kernel. The fusion kernel is derived in \cite{Drukker:2010jp}. The relation is 
\begin{align}\label{FK-action}
&	\chi_{\id}(-1/\tau) = \int [d\hat\alpha] ~S_{\id\alpha}~ \chi_\alpha(\tau)\\
&	S_{\id\alpha} =  i^{N-1} \sqrt{\det(C)} \frac{1}{|W|} \prod_{e_i} 4 \sin (\pi b \vev{ \hat a,e_i})  \sin (-\pi b^{-1} \vev{\hat a,e_i})
\end{align}
Here, $C$ is the Cartan matrix, $W$ is the Weyl group and the product above is over all the positive roots $e_i$ of $\sln$. We recollect some   facts on the representation theory of $\sln$ in Appendix \ref{app:sln} that are useful in manipulating expressions like the one above. Note that the integration measure of \eqref{FK-action} is $(N-1)$ dimensional, same as the rank of $\sln$.

We can change integration variables from $\hat\alpha$ to $\hat P$ in the fusion kernel action \eqref{FK-action}. Using $\det(C) = N$ and  $|W|=|S_N|=N!$ (see Appendix \ref{app:sln}), we have
\begin{align}\label{FK2}
	\chi_{\id}(-1/\tau) = \frac{N^{1/2}}{N!}\int [d \hat P] ~\left(\prod_{e_i} 4 \sinh (\pi b \vev{\hat P,e_i})  \sinh (\pi b^{-1} \vev{ \hat P,e_i})\right) ~ \chi_P(\tau)~. 
\end{align}
The integration limits are from $-\infty$ to $\infty$. 
It is now time to take the Schwarzian limit. As we have already seen that it emerges from the left moving character and we focus on the same for now. As before we take $\tau= i\frac{\beta_L}{2\pi}$. The Schwarzian limit is given by 
\begin{align}
	b\to 0 \implies c\to \infty~,\qquad \beta_L \sim O(c) , \qquad \hat k\equiv b^{-1} \hat P \text{~fixed}~. 
\end{align}
In this regime of parameters, equation \eqref{FK2} becomes the following
\begin{align}\label{chiFuse}
	  \chi_{\id }\left(\frac{2\pi i }{\beta_L}\right) &\approx\  \frac{(4\pi)^{N(N-1)\over 2}N^{1/2}}{N!}b^{{N^2-1}}\, e^{\beta_L {(N-1)\over 24}}\int [d\hat k] ~\left(\prod_{e_i}     \vev{\hat k,e_i}  \sinh (\pi  \vev{\hat k,e_i})\right) ~ e^{-\beta_L \frac{b^2}{2}\vev{ \hat k, \hat k}}~. 
\end{align}
The contribution from the descendants trivialises in this regime and the above integral is in the form of a sum over Boltzmann weights. Furthermore, we  recover the scaling of the central charge to be the same as the prefactor in \eqref{chiL} -- recall $\beta_L\sim c\sim b^{-2}$ and therefore
$
	\beta_L^{-({N^2-1})/{2}} \sim b^{{N^2-1}}
$ as above. 

The expression \eqref{chiFuse} can be simplified further.  First, we note that the positive roots  (now denoting them as $r_{ij}$ with $i<j$, instead of $e_i$) are given in terms of linear combinations of the simple roots $s_i$ as, see  \eqref{roots-rel}
\begin{align}
	r_{ij} =\sum_{m=i}^{j-1} s_i~. 
\end{align}
The rescaled momentum vector can also be written as  a linear combination of simple roots
\begin{align}
	\hat k= \sum_{n=1}^{N-1} \kappa_n s_n
\end{align}
Hence  Cartan-Killing form appearing in the integrand of \eqref{chiFuse} is given by
\begin{align}\label{CKForm}
	\vev{\hat k,r_{ij}} = \sum_{m=i}^{j-1} \sum_{n=1}^{N-1} \kappa_n \vev{s_n,s_m} = \sum_{m=i}^{j-1} \sum_{n=1}^{N-1}  C_{mn}\kappa_n = \sum_{m=i}^{j-1} [C  \kappa]_{m} \equiv [C  \kappa]^{(ij)}~. 
\end{align}
where, $C_{mn}$ is the Cartan matrix. In the last step we defined the partial sum of elements of the vector  $C  \kappa$ as $[C  \kappa]^{(ij)}$.  
The explicit expressions of $[C  \kappa]^{(ij)}$ can be computed systematically. For $N=2,3$ and $4$ these are
\begin{align}
	N=2:\quad &2\kappa_1~, \qquad\qquad\qquad
	N=3:\quad 2\kappa_1-\kappa_2,\, \kappa_1+\kappa_2,-\kappa_1+2\kappa_2~, \\
	N=4:\quad &2 \kappa _1-\kappa _2,\,\kappa _1+\kappa _3,\,2 \kappa _3-\kappa _2,\,\kappa _1+\kappa _2-\kappa _3,-\kappa _1+2 \kappa _2-\kappa _3,-\kappa _1+\kappa _2+\kappa _3~. \nn
\end{align}
By the same token, the norm of the momenta appearing in Boltzmann weights can be written as 
\begin{align}\label{k-norm}
	\vev{\hat k,\hat k} = \sum_{m,n=1}^{N-1} \kappa_m C_{mn} \kappa_n = \kappa ^T C  \kappa~. 
\end{align}
Using the relations \eqref{CKForm} and \eqref{k-norm}, the integral \eqref{chiFuse} can be written\footnote{The following shorthand for the product over all positive roots is used
	$
	\prod_{i<j} \equiv \prod_{i=1}^{N-1} \prod_{j=i+1}^{N}
	$. } using the components of the momentum $\kappa_m$ and the Cartan matrix of $\sln$
\begin{align}\label{spec-final}
	  \chi_{\id}\left(\frac{2\pi i }{\beta_L}\right) 
	&\approx \frac{(4\pi)^{N(N-1)\over 2}N^{1/2}}{N!}b^{N^2-1} e^{\beta_L {(N-1)\over 24}}\int [d\kappa_i] ~\left(\prod_{i<j}     [C  \kappa]^{(ij)}  \sinh (\pi [C \kappa]^{(ij)} )\right) ~ e^{-\beta_L \frac{b^2}{2}\kappa ^T C  \kappa}. 
\end{align}
Contrasting with equations \eqref{chiL} and \eqref{Z-Sch-WN}, the above formula provides  the spectral representation of higher spin Schwarzian partition function; the underlying group structure of $\sln$ is encoded via the Cartan matrix. This decomposition however is phrased using the momentum vectors characterizing the primary states as opposed to the energies. The results of the integrals appearing above and the integral in equation \eqref{dos1} (with \eqref{dos2} substituted)  are the same, \ie  they lead to the same partition function for the Schwarzian theory. However, the \textit{integrands} themselves are not amenable to a comparison; the fusion kernel \eqref{FK-action} leads to a $(N-1)$-dimensional integral, while the integral in \eqref{dos1} is one-dimensional. 
The only exception to this turns out to be  the Virasoro or $N=2$ case, the product in the integrand of \eqref{spec-final} has a single term and the integral itself is one-dimensional; upon changing variables using \eqref{h-alpha-rel} the density of states  can be directly read off, and it agrees with \eqref{N2-DoS}. 

\section{Conclusions}
\label{sec:conclusions}

In this work we studied the near-extremal limit of partition functions of higher spin fields in the BTZ black hole. It was demonstrated that the near-extremal regime contains a higher spin incarnation of the Schwarzian sector, with the spectrum generalizing in an appropriate manner. We also explored supersymmetric higher spins and found the existence of similar structures. Furthermore, we found the spectral decomposition of the partition function using the fusion kernel of $\cW_N$. 

Although we have illustrated equivalence of partition functions, it would be desirable to have a more direct approach to arrive at the BF theory from  KK reduction of 3d higher spin Chern-Simons theory around the near-horizon AdS$_2$ throat. However, this reduction needs to be formulated in a gauge-invariant manner.\footnote{In \cite{Mertens:2018fds,Gaikwad:2018dfc}, it is shown that  Chern-Simons theories in AdS$_3$ indeed KK reduces to a BF like theories. The analysis in these cases however wasn't in the near-extremal region.   } It would be of value to understand the connection of the dimensional reduction to concrete versions higher spin AdS$_2$/CFT$_1$, constructed recently in \cite{Alkalaev:2019xuv,Alkalaev:2020kut}. Also, it would be of interest to investigate scenarios with non-zero higher spin chemical potentials. These correspond to higher spin black holes in the bulk. The notion of a geometrical horizon is lost in higher spin gravity and it would be tantalizing to see how the gauge invariant notions of extremality \cite{Banados:2015tft,Henneaux:2015ywa} give rise to the Schwarzian sectors. One technical obstacle in this regard is that the one-loop partition function in the higher-spin black hole background is not known non-perturbatively. However, it is plausible that the path integral of the higher-spin Schwarzian theory can be evaluated for non-zero chemical potentials using the fermionic localization techniques of \cite{Stanford:2017thb}. This would furnish a prediction for the partition function of extremal higher spin black holes, with one-loop corrections included. 

It would also be worthwhile to explore the near-extremal limit of correlation functions of   $\Wn$ CFTs. This would enable us to evaluate more refined probes such as out-of-time-ordered correlators that serve as diagnostics for chaos. It is known that the higher spin theories violate the bound on the Lyapunov exponent \cite{Perlmutter:2016pkf,David:2017eno,Narayan:2019ove} and it would be interesting to see what is pathological from the Schwarzian perspective.

As demonstrated in \cite{Mertens:2017mtv}, the Schwarzian theory can also be seen to emerge from boundary states of Liouville theory -- the FZZT branes. A similar version of the story should also hold true when higher spin conserved currents are present; the boundary states of Toda theory need to be considered instead \cite{Drukker:2010jp}. This perspective also provides a  framework for evaluating  correlation functions \cite{Mertens:2017mtv,Mertens:2018fds} which can be utilized  to study aspects of thermalization and chaos \cite{Nayak:2019evx,Banerjee:2018kwy}. 

There has been considerable progress in understanding the path integral of JT gravity on higher genus Riemann surfaces. The key idea being formulating the path integral as a double scaled matrix model \cite{SSS}. It would be fascinating to explore the generalization for higher spin gravity, which is described by a BF theory. The results of the partition functions on the disk, provided in this work and \cite{Gonzalez:2018enk}, may provide crucial hints to proceed in this direction. 

Finally, the Schwarzian sector emerges universally as the low energy dynamics of the SYK model. In this work, we have considered higher spin generalizations of the Schwarzian theory, however, the analogue of the SYK model which would give rise to these Schwarzian actions is unclear. This offers an exciting avenue to construct such models that manifestly realize these symmetries in the IR. We note that in two-spacetime dimensions there has been constructions of supersymmetric SYK models along these lines in \cite{Peng:2018zap}.


\section*{Acknowledgements}
The author thanks Pawel Caputa, Diptarka Das, Justin David,  Arnab Kundu, Sridip Pal and, especially, Tom\'{a}\v{s} Proch\'{a}zka  for fruitful discussions. 
\appendix
\section*{Appendix}

\section{Representation theory of $\sln$ -- a few facts}
\label{app:sln}
In this appendix we collect some facts on the representation theory of $\sln$.  A good reference
for this is \cite[Appendix C]{deBoer:2013vca} from which we review certain concepts. The facts outlined here are useful for the analysis in Section \ref{sec:fusion}.

Firstly, the Cartan subalgebra of $\sln$ is $(N-1)$ dimensional. In CFT terms, a convenient basis for this is the zero-modes of the stress tensor and the higher spin currents $(L_0, W^{(s)}_0)$. The positive roots can be constructed as follows. We consider an orthonormal basis of vectors $v_i$ in $\mathbb{R}^N$. Let $\gamma$ be the sum of these vectors, $\gamma=v_i$. Next we project onto the plane orthogonal to $\gamma$ and define new vectors $u_i$ in this space as
\begin{align}
	u_i = v_i - \frac{\gamma}{N}~. 
\end{align}
These vectors are linearly dependent in the weight space of $\sln$; in particular, $\sum_i u_i=0$. The inner product is 
\begin{align}
	\vev{u_i,u_j}= \delta_{ij} - \frac{1}{N}~. 
\end{align}
The positive roots can now be constructed from the $u_i$'s. They are given by 
\begin{align}\label{positiveRoots}
r_{ij} = u_i -u_j~, \quad i<j,~ 1\leq j \leq N~. 
\end{align}
The condition $i<j$ indicates the upper triangular matrix for $r_{ij}$. Therefore there are $N(N-1)/2$ positive roots. The simple roots can be constructed from these as 
\begin{align}
	s_i = r_{i,i+1} = u_i - u_{i+1}~. 
\end{align}
and there are $(N-1)$ of these. The roots are invariant under the action of the symmetric group $S_N$. This is the Weyl group of $\sln$. The Weyl vector, $\hat\rho$,  is the sum of all positive roots
\begin{align}
	\hat\rho = \frac{1}{2}\sum_{j=1}^N  \sum_{i=1}^j r_{ij}~. 
\end{align}
We also have the relation
\begin{align}\label{roots-rel}
	r_{ij}= r_{i,i+1}+r_{i+1,i+2}+\cdots + r_{j-1,j} =\sum_{m=i}^{j-1} s_i~. 
\end{align}
The first equality follows from the fact the positive roots can themselves be written as differences of a pair of vectors, \cref{positiveRoots}. The second equality uses the definition of simple roots $s_i =r_{i,i+1}$. 
Finally, the Cartan matrix is defined by the following inner product of simple roots and takes a band-diagonal form
\begin{align}
	C_{ij} = 2 \frac{\vev{s_i,s_j}}{\vev{s_j,s_j}} = \vev{s_i,s_j} = \begin{cases}
		2,~ ~~\,i=j ~,\\
		-1,~ i=j\pm 1 ~. 
	\end{cases}
\end{align} 



\begin{small}
	\providecommand{\href}[2]{#2}	
\end{small}

\end{document}